\documentclass[apj,revtex4,numberedappendix]{emulateapj}
\usepackage{graphics,graphicx,subfigure,float,color,amsmath,natbib}
\usepackage{hyperref}
\usepackage[dvipsnames]{xcolor}
\hypersetup{urlcolor=blue} 
 
\setcitestyle{notesep={ }}

\newcommand{\CH}[1]{\colhead{#1}}
\newcommand\W{{$\lambda$}}

\defcitealias{berg16}{B16}
\defcitealias{berg19a}{B19}

\begin{document}

\shortauthors{Berg et al.}
\title{Intense \ion{C}{4} and \ion{He}{2} Emission in $z\sim0$ Galaxies: \\
Probing High-Energy Ionizing Photons\footnote{
Based on observations made with the NASA/ESA Hubble Space Telescope,
obtained from the Data Archive at the Space Telescope Science Institute, which
is operated by the Association of Universities for Research in Astronomy, Inc.,
under NASA contract NAS 5-26555.}}

\author{Danielle A. Berg\altaffilmark{1}, 
	    John Chisholm\altaffilmark{2},
	    Dawn K. Erb\altaffilmark{3},
	    Richard Pogge\altaffilmark{1,4},
	    Alaina Henry\altaffilmark{5},
	    Grace M. Olivier\altaffilmark{1}
	    }
\altaffiltext{1}{Department of Astronomy, The Ohio State University, 140 W 18th Avenue, Columbus, OH 43210, USA}
\altaffiltext{2}{Department of Astronomy \& Astrophysics, UC Santa Cruz, 156 High Street, Santa Cruz, CA, USA}
\altaffiltext{3}{Center for Gravitation, Cosmology and Astrophysics, Department of Physics, University of Wisconsin Milwaukee, 3135 N Maryland Ave., Milwaukee, WI 53211, USA}
\altaffiltext{4}{Center for Cosmology \& AstroParticle Physics, The Ohio State University, 191 W Woodruff Avenue, Columbus, OH 43210}
\altaffiltext{5}{Space Telescope Science Institute, 3700 San Martin Drive, Baltimore, MD, 21218, USA}


\begin{abstract}
In the last few years, prominent high-ionization nebular emission lines (i.e., \ion{O}{3}], 
\ion{C}{3}], \ion{C}{4}, \ion{He}{2}) have been observed in the deep UV spectra of $z\sim5-7$ 
galaxies, indicating that extreme radiation fields characterize reionization-era systems. 
These lines have been linked to the leakage of Lyman continuum photons 
(necessary for reionization) both theoretically and observationally. 
Consequently, high-ionization UV emission lines present our best probe to detect and characterize 
the most distant galaxies that we will observe in the coming years, and are key to understanding 
the sources of reionization, yet the physics governing their production is poorly understood.
Here we present recent high-resolution {\it Hubble Space Telescope} spectra of two 
nearby extreme UV emission-line galaxies, J104457 and J141851.
We report the first observations of intense nebular \ion{He}{2} and double-peaked, 
resonantly-scattered \ion{C}{4} emission, a combination that suggests these galaxies 
both produce and transmit a significant number of very high-energy ionizing photons 
($E > 47.89$ eV) through relatively low column densities of high-ionization gas. 
This suggests that, in addition to photons at the H-ionizing edge,
the very hard ionizing photons that escape from these galaxies may provide a secondary source 
of ionization that is currently unconstrained observationally. 
Simultaneous radiative transfer models of Ly$\alpha$ and \ion{C}{4} are needed to 
understand how ionizing radiation is transmitted through both low- and high-ionization gas.
Future rest-frame FUV observations of galaxies within the epoch of reionization using the 
James Webb Space Telescope ({\it JWST}) or extremely large telescopes (ELTs) will allow us to 
constrain the escape of helium-ionizing photons and 
provide an estimate for their contribution to the reionization budget. 
\end{abstract}

\keywords{galaxies: abundances - galaxies: evolution}


\section{Introduction}\label{sec1}
The 21st century of astronomy has been marked by deep imaging surveys with the
{\it Hubble Space Telescope} (HST) that have opened new windows onto the high-redshift
universe, unveiling thousands of $z>6$ galaxies
\citep[e.g.,][]{bouwens15,finkelstein15,livermore17,atek18,oesch18}.
From these studies, and the numerous sources discovered, a general consensus
has emerged that low-mass galaxies host a substantial fraction of the star formation in 
the high-redshift universe and are likely the key contributors to reionization 
\citep[e.g.,][]{wise14,robertson15,madau15,stanway16}. 
Much work has been done to characterize these reionization epoch systems, revealing compact,
metal-poor, low-mass galaxies with blue UV continuum slopes and large specific star 
formation rates \citep[sSFRs; e.g.,][]{stark16}. 
Additionally, deep restframe UV spectra of $z\sim5-7$ galaxies have revealed prominent 
high-ionization nebular emission lines (i.e., \ion{O}{3}], \ion{C}{3}], \ion{C}{4}, \ion{He}{2}) 
indicating that extreme radiation fields characterize reionization-era galaxies 
\citep{sobral15,stark15,stark16,mainali17,mainali18}, and may be more common at 
high redshifts \citep[see, also,][]{smit14,smit15}.\looseness=-2

Currently, few direct detections of the H-ionizing Lyman continuum photons (LyC; \W $< 912$ \AA)
exist at high redshifts \citep[e.g., $z>3$,][]{debarros16,shapley16,vanzella18,steidel18}, 
owing to their faintness and the increasing neutral intergalactic medium (IGM) 
attenuation with redshift \citep[e.g.,][]{inoue14}.
Alternatively, studies of local galaxies with similar properties to observed 
reionization-era galaxies have lead to the proposal of several LyC-emitter indicators. \looseness=-2

Ly$\alpha$ is the strongest feature in the rest-frame far-UV (FUV) emission-line spectra 
of galaxies, and is commonly used as an indirect indicator of leaking ionizing radiation. 
Recent radiative transfer simulations have demonstrated that low column densities of 
low-ionization (\ion{H}{1}) gas lead to multi-peaked Ly$\alpha$ emission profiles 
\citep[e.g.,][]{verhamme06,verhamme15,gronke16,dijkstra06,dijkstra14}. 
In these models, smaller velocity separations between the two emission peaks correlate with 
lower \ion{H}{1} column densities within the galaxy through which LyC photons can escape.
Observationally, all LyC emitting galaxies have strong double-peaked Ly$\alpha$ emission profiles 
\citep[e.g.,][]{verhamme17}, which are produced in media that either have low \ion{H}{1} column 
densities or inhomogeneous covering fractions of gas through which ionizing radiation could escape.
Strong correlations have been reported between the LyC escape fraction of local leakers 
with the shape of the Ly$\alpha$ emission line \citep{izotov18} and with 
the fraction of Ly$\alpha$ photons that escape galaxies \citep{chisholm18a}. 
In other words, Ly$\alpha$ emission is our most reliable indirect indicator of LyC escape. 
However, blueshifted Ly$\alpha$ emission is also attenuated by the IGM from galaxies within 
the epoch of reionizaiton and so cannot diagnose LyC escape at high-$z$. \looseness=-2

Alternatively, rest-frame high-ionization nebular emission lines have been 
proposed as probes of reionization-era systems. 
Indeed, strong high-ionization nebular emission, which indicates a large production 
of high-energy ionizing photons, has been linked to the leakage of LyC photons both 
theoretically \citep[e.g.,][]{nakajima14} and observationally 
\citep[e.g.,][]{izotov16, chisholm18a}. 
Specifically, the large ratios of high- to low-ionization emission that have been 
observed from all galaxies emitting LyC,
have been previously used to argue for fully-ionized interstellar media (i.e., are density bounded) 
or optically-thin tunnels by which ionizing photons can escape 
\citep[e.g.,][]{jaskot13, nakajima14, rivera-thorsen15, gazagnes18}.
This idea is further supported by the low covering fractions of low-ionization absorption lines
observed in many high-ionization galaxies \citep[e.g.,][]{heckman11,alexandroff15,henry15,erb16,chisholm18a}. \looseness=-2

LyC indicators are intimately intertwined at high-redshift, 
where high-ionization UV emission lines have been detected in five of the 13 known Ly$\alpha$ 
emitters at $z>7$.
These observations suggest the extreme radiation fields more commonplace at high redshift 
may aid the transmission of Ly$\alpha$ photons, and consequently ionizing photons 
as they must travel through the same gas \citep{mainali18}. \looseness=-2

Therefore, high-ionization nebular UV emission lines provide a valuable tool to identify
potential galaxies with high transmission of ionizing photons and characterize the most distant
galaxies that the next generation of telescopes will observe.
Yet, the physics governing the production of high-ionization nebular emission lines and their 
link to Ly$\alpha$ and LyC escape are poorly understood. 
Of these lines, \ion{C}{4} emission may be especially informative as it is a resonant line, 
meaning it is easily scattered by the gas surrounding galaxies, and so, similar to Ly$\alpha$, 
is very sensitive to column density. 
While the Ly$\alpha$ profile traces neutral gas, \ion{C}{4} emission probes the density 
of the high-ionization gas through which very high energy ionizing photons may escape.
\ion{C}{4} emission, therefore, provides the rare opportunity to quantify \textit{both} 
the production and transmission of high-energy ionizing photons.
Here we present high-resolution {\it HST} spectra of the intense, high-ionization 
UV \ion{C}{4} and \ion{He}{2} nebular emission in two $z\sim0$ galaxies with 
properties typical of reionization-era sources. \looseness=-2


\begin{deluxetable}{rlcc}
\tablecaption{Extreme UV Emission-Line Galaxy Properties}
\tablehead{
\multicolumn{1}{r}{Property}  & \multicolumn{1}{l}{Units}	& \CH{J104457} 		& \CH{J141851}  }
\startdata	
\multicolumn{4}{c}{\bf Adopted from Archival Sources:} \\
Reference				&					    & \citetalias{berg16}       & \citetalias{berg19a} \\
R.A.					& J2000				    & \ \ \ 10:44:57.79	        & \ \ 14:18:51.13	\\
Decl.					& J2000				    & $+$03:53:13.15	        & $+$21:02:39.74	\\
$z$					    &					    & 0.013	                    & 0.009         	\\
log $M_\star$			& M$_\odot$			    & 6.80		                & 6.63          	\\
log SFR				    & M$_\odot$ yr$^{-1}$	& $-0.85$		            & $-1.16$       	\\
log sSFR				& yr$^{-1}$			    & $-7.64$		            & $-7.79$         	\\
$E_(B-V)$				& mag.				    & 0.077				        & 0.140		    	\\
$12+$log(O/H)			& dex $\mid Z_\odot$	& $7.45 \mid 0.058$ 	    & $7.54 \mid 0.071$	\\
log $U$					&					    & $-1.77$				    & $-2.42$			\\
\\
\multicolumn{4}{c}{\bf Derived from the UV COS G160M Spectra:} \\
EW$_{\rm OIII]}$		& \AA\			        & $-2.89,-6.17$				& $-1.68,-4.78$		\\
EW$_{\rm CIV}$		    & \AA\			        & $-6.71,-2.83$				& $-1.78,-1.43$		\\
EW$_{\rm HeII}$		    & \AA\			        & $-2.75$					& $-2.82$			\\
\ion{O}{3}] $\Delta V$ 	& km s$^{-1}$		    & $87.3$					& $64.7$			\\
\ion{C}{4} $\Delta V$	& km s$^{-1}$		    & \color{blue}$106.7$\color{black}, \color{red}$86.2$\color{black}				
                                                & \color{blue}$101.8$\color{black}, \color{red}$78.6$\color{black}  \\
\ion{He}{2} $\Delta V$	& km s$^{-1}$		    & $110.5$					& $85.2$			\\
CIV $V_{peak}^{blue}$	& km s$^{-1}$		    & $-66.9$					& $-24.2$			\\
CIV $V_{peak}^{red}$	& km s$^{-1}$		    & $+41.7$					& $+77.7$			\\
CIV $V_{sep.}$			& km s$^{-1}$		    & $108.6$					& $103.9$			\\
L$_{1500}$			    & erg s$^{-1}$ Hz$^{-1}$ & $3.48\times10^{27}$	    & $1.23\times10^{27}$ \\
\\
\multicolumn{4}{c}{$f_\gamma = 1.0$, $f_{esc} = 0.0$:} \\
Q$_0$           		& s$^{-1}$              & $4.70\times10^{52}$       & $1.88\times10^{52}$ \\
Q$_1$           		& s$^{-1}$              & $3.75\times10^{51}$       & $1.38\times10^{51}$ \\
Q$_2$           		& s$^{-1}$              & $3.87\times10^{50}$       & $1.61\times10^{50}$ \\
log $\xi_{\rm ion}^0$	& Hz erg$^{-1}$         & 25.13					    & 25.18		          \\	
\\
\multicolumn{4}{c}{$f_\gamma = 0.8$, $f_{esc} = 0.2$:} \\
Q$_0$           		& s$^{-1}$              & $5.88\times10^{52}$       & $2.36\times10^{52}$ \\
Q$_1$           		& s$^{-1}$              & $4.69\times10^{51}$       & $1.73\times10^{51}$ \\
Q$_2$           		& s$^{-1}$              & $4.84\times10^{50}$       & $2.01\times10^{50}$ \\
log $\xi_{\rm ion}^{0}$	& Hz erg$^{-1}$         & 25.23				        & 25.28	              \\
log $\xi_{\rm ion}^{0,1,2}$	& Hz erg$^{-1}$     & 25.26				        & 25.31	      
\enddata	
\tablecomments{ 
Properties of the extreme UV emission-line galaxies presented here.
The top portion of the table lists properties previously reported by
\citet{berg16} for J104457 and \citet{berg19a} for J141851.
The R.A., Decl., redshift, total stellar masses, SFRs, and sSFRs were adopted from the SDSS MPA-JHU DR8 
catalog\footnote{Data catalogues are available from \url{http://www.sdss3.org/dr10/spectro/galaxy_mpajhu.php}.
The Max Plank institute for Astrophysics/John Hopkins University(MPA/JHU) SDSS data base was produced by a 
collaboration of researchers(currently or formerly) from the MPA and the JHU. 
The team is made up of Stephane Charlot (IAP), Guinevere Kauffmann and Simon White (MPA),
Tim Heckman (JHU), Christy Tremonti (U. Wisconsin-Madison $-$ formerly JHU) and Jarle 
Brinchmann (Leiden University $-$ formerly MPA).},
whereas $E(B-V)$, $12+$log(O/H), and log $U$, were measured from the SDSS optical spectra.
The bottom portion of the table lists the properties derived in this work from the UV
HST/COS G160M spectra (see Figure~2).
Equivalent widths are listed for \ion{C}{4} \W\W1548,1550, \ion{O}{3} \W\W1661,1666, and \ion{He}{2} \W1640.
We list velocity widths, $\Delta V$, of \ion{O}{3}] \W1666 and \ion{He}{2} \W1640,
and both blue and red components of the \ion{C}{4} \W1548 profiles.
Since the individual \ion{C}{4} lines present as doublets, we list the 
$V_{peak}^{blue}$, $V_{peak}^{red}$, and $V_{sep.}$ measured for the \ion{C}{4} \W1548 line.
The \ion{H}{1} (\ion{He}{1}, \ion{He}{2}) ionizing photons, Q$_0$ (Q$_1$, Q$_2$) 
are derived from the H$\beta$ (\ion{He}{1} \W4471, \ion{He}{2} \W4686) flux.
The UV luminosity density is derived from the UV continuum at 1500\AA.
Together, these values are used to determine log $\xi_{\rm ion}^0$ (see Section 3.2). \looseness=-2}
\label{tbl1}
\end{deluxetable}
   

\begin{figure}
\begin{center}
	\includegraphics[scale=0.27,trim=0mm 0mm 0mm 0mm,clip]{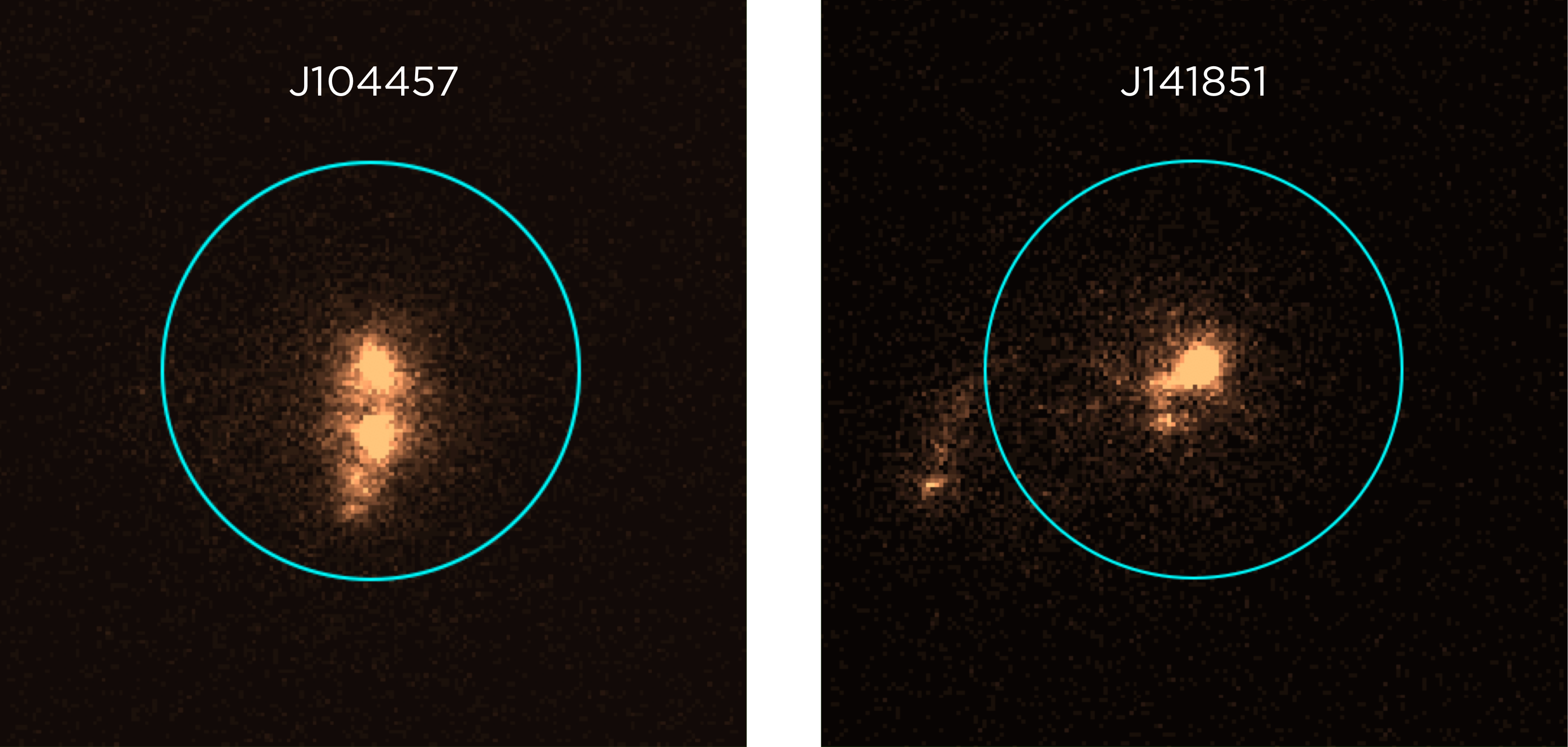}
\caption{COS NUV acquisition images demonstrating the compact UV morphologies of our galaxies
relative to the 2.\arcsec5 COS aperture (blue circle).}
\end{center}
\end{figure}


\section{Observations of $z\sim0$ Reionization Era Analogues}
\subsection{Extreme UV Emission-Line Galaxies}
Recent {\it HST}/COS spectral campaigns have revealed a sample of nearby galaxies with emission-line 
features more analogous to reionization-era systems than their local family of galaxies. 
In particular, studies targeting high-ionization UV emission lines in nearby,
metal-poor (12+log(O/H)$\lesssim8.0$) star-forming galaxies have observed
22 objects with significant nebular emission from \ion{C}{4} 
\W\W1548,1550 or \ion{He}{2} \W1640 \citep{berg16,senchyna17,berg19a,senchyna19}.
Of these, J104457 and J141851, have the largest reported UV \ion{C}{4} EW 
\citep[][\llap, hereafter, B16]{berg16} and \ion{He}{2} EW \citep[][\llap, hereafter, B19]{berg19a},
respectively, amongst local star-forming galaxies.\footnote{Note that the \ion{C}{4} 
and \ion{He}{2} EWs reported in Table~1 are from the higher-resolution spectra 
presented here and are lower than (but within the errors of) the EWs 
originally reported in \citetalias{berg16} and \citetalias{berg19a}.} \looseness=-2

Here we present high-resolution {\it HST}/COS G160M spectra of two
extreme UV emission-line galaxies, J104457 and J141851.
These galaxies are shown in Figure~1 and their properties, many of which liken 
them to reionization-era systems, are listed in Table~\ref{tbl1}.
Despite the strong \ion{C}{4} and \ion{He}{2} emission detected in these galaxies,
\citetalias{berg16} and \citetalias{berg19a} find that their relative UV and optical emission-line 
ratios are typical of intensely star-forming galaxies, and not AGN.
For example, the strong \ion{O}{3}] \W\W1661,1666 emission shown in Figure~2
has not been observed for narrow-line AGN \citep[e.g.,][]{hainline11}, where
the very hard radiation fields producing strong \ion{He}{2} recombination lines 
also weaken \ion{O}{3}] by triply-ionizing O instead.

Stellar population synthesis (SPS) models that include very massive stars (M$\sim300$M$_\odot$) 
or binaries can reproduce the observed \ion{C}{4}, \ion{O}{3}], and \ion{C}{3}] emission,
but cannot explain \ion{He}{2} \citep[e.g.,][]{berg18,stanway19}.
Relative to the \ion{He}{2} emission, 
which requires the largest ionization energy of these lines ($>54$ eV),
standard SPS models underpredict, by an order of magnitude, the number of ionizing photons 
produced beyond the He$^+$-ionizing edge relative to the H-ionizing flux. 
Alternatively, \citet{schaerer19} suggest high-mass X-ray binaries as the source of 
\ion{He}{2} ionizing photons in star-forming galaxies, while \citet{senchyna19}
propose the presence of strong, nebular \ion{C}{4} emission as an indicator of 
ionizing photons from very low-metallicity ($Z < 0.1\ Z_\odot$) massive stars.
While we may rule out AGN, it is still unclear what is powering this unique suite of relative 
line strengths of \ion{C}{4}, \ion{He}{2}, \ion{O}{3}], and \ion{C}{3}]. \looseness=-2

Despite the serendipitous success of recent surveys, it is important to note that 
very strong high-ionization UV emission lines are atypical of $z\sim0-3$ galaxy samples.
This is due, in part, to the fact that nebular \ion{He}{2} and \ion{C}{4} are only found 
at low metallicities (12+log(O/H)$\lesssim8.0$) and few UV spectroscopic campaigns 
have targeted the 1500--1700 \AA\ range of metal-poor galaxies.
Even so, the low-metallicity targeted surveys considered here only have a 
$\sim25$\% detection rate of both nebular \ion{He}{2} and \ion{C}{4}.
These extreme nebular features are far from ubiquitous even amongst 
high-ionization, very metal-poor galaxies \citep[e.g.,][]{senchyna19}. \looseness=-2


\begin{figure*}
\begin{center}
\begin{tabular}{c}
	\includegraphics[scale=0.25,trim=0mm 0mm 0mm 0mm,clip]{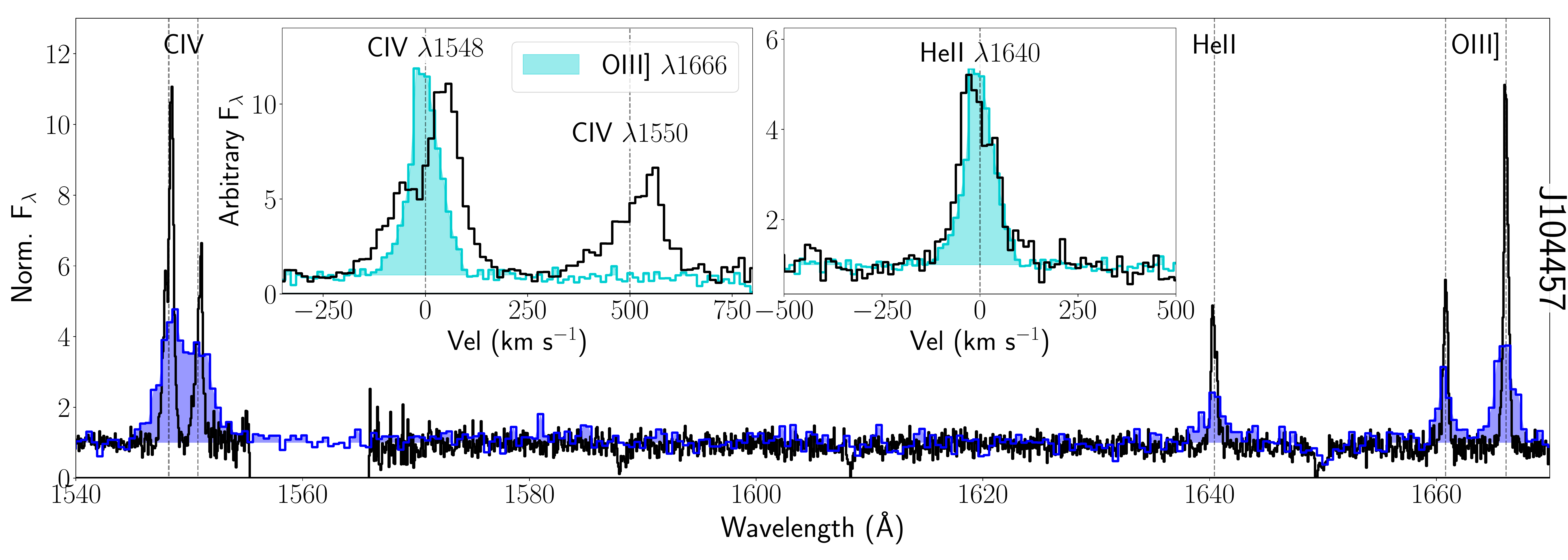}  \\
	\includegraphics[scale=0.25,trim=0mm 0mm 0mm 0mm,clip]{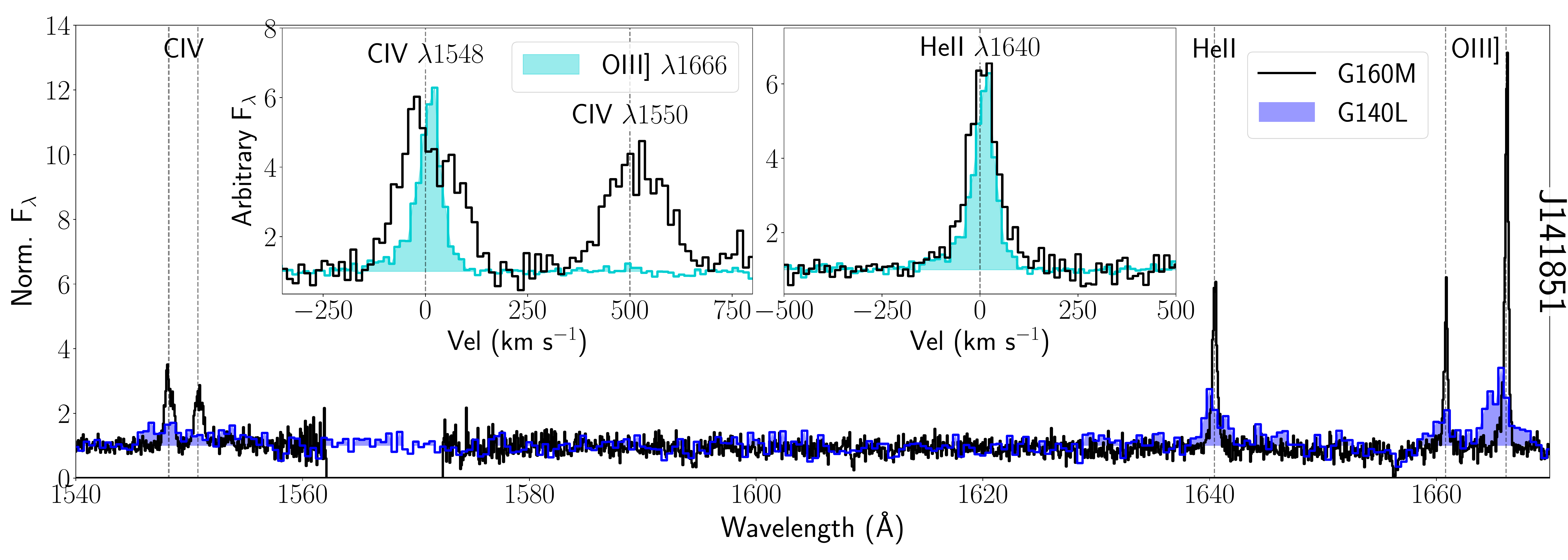}
\end{tabular}	
\caption{
Comparison of the low-resolution {\it HST}/COS G140L spectra (filled blue histogram)
with the higher resolution {\it HST}/COS G160M spectra (black line)
for J104557 ({\it top}; \citetalias{berg16}) and J141851 ({\it bottom}; \citetalias{berg19a}).
The {\it left} inset spectra show the velocity profiles of the resonantly-scattered \ion{C}{4} 
\W\W1548,1550 doublet (black line) versus a nebular \ion{O}{3}] emission line (filled green line), 
demonstrating that the individual \ion{C}{4} lines are broadened and appear to be double peaked.
In the {\it right} inset panels, however, we plot the \ion{He}{2} \W1640 emission feature (black line) versus
the nebular \ion{O}{3}] \W1666 emission line (filled green line) and find it to be narrow and nebular in origin.}
\end{center}
\end{figure*}


\subsection{HST/COS Spectra}
Strong \ion{C}{4} and \ion{He}{2} emission were originally observed for
J104457 and J141851 in their low-resolution {\it HST}/COS G140L spectra.
Based on these detections, 
we obtained high-resolution {\it HST}/COS G160M observations with program HST-GO-15465.
Our observing strategy was similar to \citetalias{berg19a}, 
but used the G160M grating at a central wavelength of 1589\AA,
and science exposures of 6439 and 12374s for J104457 and J141851, respectively.
In order to improve the signal-to-noise, we binned the spectra by 6 native COS pixels
such that $\Delta v = 13.1\ {\rm km\ s}^{-1}$ and the emission line FWHMs are sampled 
by more than 4 pixels (see Table~1). 

The new G160M spectra for J104457 and J141851 are compared to
the low-resolution G140L spectra in Figure~2.
The higher spectral resolution and S/N of the G160M spectra allow us to investigate the 
kinematic properties of the most intense high-ionization UV \ion{C}{4} and \ion{He}{2} 
emission features in local galaxies for the first time.
We reserve the detailed modeling analysis of the stellar continuum, ISM features, and 
nebular emission for a future paper \citep{olivier19}. \looseness=-2

\subsection{Resonantly Scattered \ion{C}{4} Emission}
The \ion{C}{4} \W\W1548,1550 spectral feature is challenging to interpret due to complex composite
profiles with possible contributions from narrow nebular emission, broad stellar emission, stellar 
photospheric absorption, and interstellar medium absorption \citep{leitherer11}.
Further, \ion{C}{4} is a resonant transition such that high-ionization gas
scatters the photons.
The resulting profiles of the reprocessed \ion{C}{4} emission for this work are plotted 
in velocity space in the left inset panels of Figure~2.
For comparison, we plot an \ion{O}{3}] emission line to represent the width of nebular emission.
We find the resonantly-scattered \ion{C}{4} emission lines to be visibly broader, 
with profile widths nearly twice that of \ion{O}{3}].
However, the strength and pure emission of the \ion{C}{4} profiles,
and therefore the absence of ISM absorption, are evidence of 
channels of relatively low column density high-ionization gas along the line of sight 
that allows its transmission. \looseness=-2

The \ion{C}{4} \W1548 features also appear to be double-peaked
\citep[see, also, the double-peaked \ion{Si}{4} emission in][]{jaskot17}.
This double-peaked profile is a classic signature of resonantly scattered emission 
and is found in the Ly$\alpha$ emission profiles of LyC emitting galaxies \citep[e.g.,][]{verhamme17}
and predicted LyC emitters \citep[e.g.,][]{henry15,henry18}.
We characterize the \ion{C}{4} \W1548 profiles of our galaxies by fitting a double 
Gaussian with unconstrained wavelengths and line widths.
From the fits we measure the maximum peak both blueward ($V_{blue}^{peak}$)
and redward ($V_{red}^{peak}$) of the systemic redshift and determine
the peak separation as $V_{sep.} = V_{blue}^{peak} - V_{red}^{peak}$.
J104457 has a blue-peak dominant profile, with a peak separation of $V_{sep.}\sim109$ km s$^{-1}$,
whereas J141851 has a red-peak dominant profile, but with a similar peak separation of 
$V_{sep.}\sim104$ km s$^{-1}$. \looseness=-2

The \ion{C}{4} \W1548 profiles for both galaxies are well fit by the 
double Gaussians assumed here (see Figure~3).
Because \W1550 is the weaker line of the \ion{C}{4} doublet, its 
profiles are noisier and more difficult to freely characterize.
Instead, we use the theoretical relative emissivities, given the measured temperature and density, 
to scale the \ion{C}{4} \W1548 fit to the \W1550 profile.
This method successfully reproduces the \W1550 line profile for J104457 and 
J141851, where the excess of red emission seen in J141851 is within the error. 
Given the fact that the bluer \W1548 line experiences a larger optical depth
to resonant scattering (with relative oscillator strengths of $f_{1548}/f_{1550}\sim2$), this exercise supports
the low column density nature of high-ionization gas in these galaxies. \looseness=-2

Radiative transfer models of Ly$\alpha$ resonance emission indicate that 
small peak separations correlate to large escape fractions \citep[e.g.,][]{verhamme15}, 
due to low average column densities of gas that may be uniform or include lower density channels.
Analogously, we expect the \ion{C}{4} \W1548 peak separations to decrease 
with lower column densities of high-ionization gas, which also 
lowers the opacity to high-energy ionizing photons. 
The very strong nebular \ion{C}{4} emission observed in J104457 and J141851
suggests that these galaxies have low column densities of high-ionization gas that may allow
large escape fractions of high-energy ionizing photons.
In this work we have measured, for the first time, both \ion{C}{4} \W\W1548,1550 emission 
lines to have consistent double-peaked resonantly-scattered profiles with extremely narrow 
peak separations;
these fits, reported in Table~1, will be useful to constrain future \ion{C}{4} radiative transfer modeling. \looseness=-2


\subsection{Nebular \ion{He}{2} Emission}\label{sec:he2}
In the right inset panels of Figure~2 we plot the velocity structure of 
the \ion{He}{2} emission versus the nebular \ion{O}{3}] \W1666 profile.
For both galaxies, the \ion{He}{2} profile widths are $\sim20$ km s$^{-1}$ wider than the \ion{O}{3}] 
widths (see Table~1), possibly due to an unresolved stellar component.
However, visually, the \ion{He}{2} emission appears consistent with the \ion{O}{3}] profiles,
indicating the \ion{He}{2} emission is dominated by nebular emission and that very hard 
ionizing sources must be present, 
comparable to the known 
UV emission-line galaxies at high-redshift. 
\citet{stark15} measure a \ion{He}{2} (upper limit) flux that is roughly 
27\% of the \ion{C}{4} emission in a $z\sim7$ galaxy, whereas we observe 
\ion{He}{2}/\ion{C}{4} flux at the levels of 22\% and 86\%, respectively. 
Given these unprecedented levels of relative \ion{He}{2} emission 
\citep[see, also,][for a $z\sim2$ galaxy with \ion{He}{2}/\ion{C}{4}$=0.46$]{berg18}, 
we could be witnessing hard radiation from X-ray binaries
in very low-mass galaxies; future X-ray observations will help constrain the 
nature of these ionizing sources. \looseness=-2

UV \ion{He}{2} can go undetected in known optical \ion{He}{2}
emitters due to extinction.
Six of the 13 galaxies from \citetalias{berg16} and \citetalias{berg19a} with \ion{He}{2} 
\W4686 detections in their optical SDSS spectra do not have 3$\sigma$
\ion{He}{2} \W1640 detections, where the missing flux can be accounted for by dust attenuation.
However, the large \ion{He}{2} \W1640/\W4686 ratios observed 
for J104457 and J141851 confirm low dust attenuation of $E(B-V)\sim0.1$
\citep[assuming an intrinsic ratio of \W1640/\W4686=6.96;][]{dopita03}.
Even when very strong nebular \ion{He}{2} \W1640 is detected, such as in
I~Zw~18, \ion{C}{4} is often seen in absorption \citep[e.g.,][]{lebouteiller13,senchyna19}
likely due to higher column densities of high-ionization gas.
Therefore, while optical \ion{He}{2} \W4686 nebular emission is not uncommon in metal-poor 
blue compact dwarf galaxies \citep[BCDs; e.g.,][]{guseva00,thuan05,shirazi12}, 
the combination of strong UV \ion{C}{4} and \ion{He}{2} emission observed in our galaxies 
is much more rare, perhaps indicating unique physical conditions that allow their detection. \looseness=-2


\begin{figure*}
\begin{center}
\begin{tabular}{c}
	\includegraphics[scale=0.25,trim=5mm 0mm 0mm 0mm,clip]{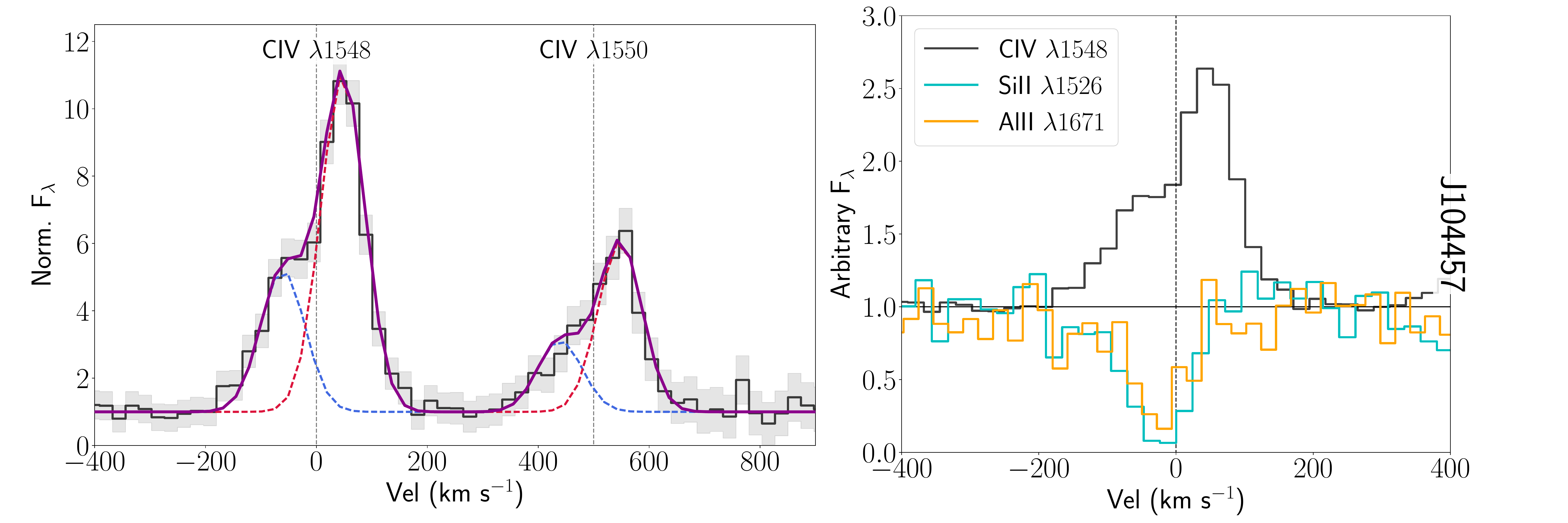}  \\
	\includegraphics[scale=0.25,trim=5mm 0mm 0mm 0mm,clip]{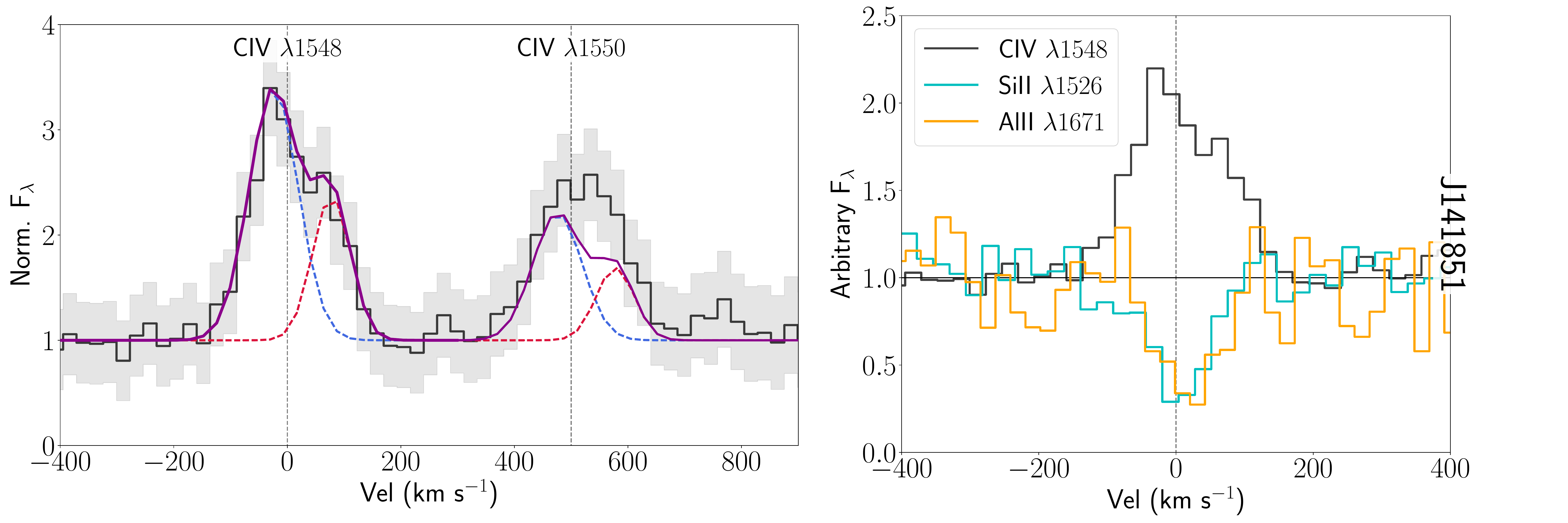}
\end{tabular}	
\caption{
{\it Left column:} We fit the observed \ion{C}{4} \W1548 profiles (black lines) of J104457 (top) and J141851 (bottom) with double Gaussian profiles 
(purple line, with blue and red components) and use the theoretical ratio of \ion{C}{4} \W1548/\W1550 to scale the \W1550 fit. 
The \W1550 feature is only marginally fit within the errors (gray) for J141851, but is a good fit for J104457.
{\it Right column:} We compare the velocity structures of the \ion{C}{4} \W1548 profiles to low-ionization absorption lines.
In both galaxy cases, the profiles all have centers near the systemic velocity, indicating little-to-no outflows.
Note, however, the \ion{Si}{2} and \ion{Al}{2} profiles may be saturated for J104457.}
\end{center}
\end{figure*}



\section{Escape of High Energy Photons}\label{sec2}

The observed strong nebular \ion{C}{4} and \ion{He}{2} emission require an especially hard 
ionizing radiation field ($> 47.9$ and 54.4 eV respectively), indicating the production 
of an unusually large number of high-energy photons.
We discuss below how the strong \ion{C}{4} emission that is resonantly scattering 
out of these galaxies indicates physical conditions that may also allow high-energy photons to escape. \looseness=-2


\subsection{H and He Opacities}
The photoionization cross-sections of H and He are important factors in 
determining the fraction of ionizing photons that can escape. 
At the energies needed to ionize C$^{+2}$ to C$^{+3}$ \citep[47.888 eV,][]{draine11}
and He$^{+}$ to He$^{+2}$ (54.418 eV), the H cross-sections are only 
$\sigma_{\rm H}({\rm C}^{+3}) = 1.4425\times10^{-19} {\rm cm}^2$ and 
$\sigma_{\rm H}({\rm He}^{+2}) = 9.834\times10^{-20} {\rm cm}^2$,
or less than 2.29\% of the maximum cross-section at 13.6 eV.
Neutral H gas is essentially optically-thin to high-enery ionizing photons.
Instead, at the wavelengths corresponding to C$^{+3}$ (\W$<258.9$\AA) and 
He$^{+2}$ (\W$<227.8$\AA) ionization, the \ion{He}{1} bound-free opacity dominates.

To understand the effects of the He opacity on the escape of high-energy photons,
we refer to the circumgalactic medium transmission models of \citet{mccandliss17},
which include the neutral H, neutral He, and singly ionized He optical depths
as products of the column densities and cross-sections for each species.
Specifically, in their Figure~3 these authors compute the LyC transmission functions 
for a range of \ion{H}{1} column densities that span the transition from optically 
thin to thick at the Lyman edge.
For a moderate column density of log(N$_{\rm HI}/{\rm cm}^{-2}) = 18.5$
the transmission of \ion{C}{4} ionizing photons (\W$<258.9$\AA) is higher
than $\sim35$\%, and increases exponentially with decreasing column density.
At this column density, the transmission of \ion{He}{2} ionizing photons 
(\W$<227.8$\AA) is $\gtrsim45$\%.
These transmission models show that the integrated escape fraction of 
ionizing photons can be as large as 27\% even when the Lyman
edge appears opaque for $17.9<$log(N$_{\rm HI}/{\rm cm}^{-2})<18.5$ 
\citep[see Figure 4 of][]{mccandliss17}.

Therefore, given the relatively low column densities of high-ionization gas in our
galaxies, as demonstrated by their strong \ion{C}{4} emission and lack of ISM absorption, 
the \citet{mccandliss17} models suggest that a significant fraction of high-energy 
ionizing photons can escape, even if \ion{H}{1} column densities are large
(no observations of N$_{\rm HI}$ exist for these galaxies).
Consequently, the combination of strong \ion{C}{4} and \ion{He}{2} emission could be 
good indicators for the production {\it and} escape of the high-energy ionizing photons 
that could contribute to reionization in the early universe. \looseness = -2


\subsection{H-Ionizing Photon Production Efficiency ($\xi_{\rm ion}$)}
Next, we consider the ionizing photon production efficiency, $\xi_{\rm ion}$.
Following the method of \citet{shivaei18}, we used the ratio of the H ionizing photon 
luminosity to the non-ionizing UV continuum luminosity density:
\begin{equation}
    \xi_{\rm ion} = \frac{Q_0}{L_{\rm UV}}\ ({\rm Hz\ erg}^{-1}).
\end{equation}
Here, $L_{\rm UV}$ is determined from the continuum flux of the G160M spectra at 1500\AA.
The continuum flux was dereddened using the \citet{calzetti00} extinction law, assuming $R_V = 4.05$
and $E_S(B-V) = 0.44\times E(B-V)$, where $E(B-V)$ is determined from the optical spectrum.
The H ionizing photon luminosity, $Q_0$, is then estimated from the dereddened SDSS H$\beta$ flux as
\begin{equation}
    Q_0 \approx \frac{4\pi D_l^2 I_{{\rm H}\beta}}{h\nu_{{\rm H}\beta}f_{\gamma}} \frac{\alpha_{\rm B}}{\alpha_{{\rm H}\beta}^{eff}}.
\end{equation}
Here, $\alpha_{\rm B}$ and $\alpha_{{\rm H}\beta}^{eff}$ are the total and effective H$\beta$ 
case-B H recombination coefficients calculated for the electron temperatures of our targets \citep{pequignot91}. 
The fraction of ionizing photons absorbed by the gas, $f_{\gamma} = 1.0 - f_{esc}$, is equal to 1.0
for a standard ionization-bounded model, providing a lower limit on the number
of \ion{H}{1} ionizing photons produced.

In addition to the H ionizing photons, we also determine the number of 
\ion{He}{1} and \ion{He}{2} ionizing photons, $Q_1$ and $Q_2$, respectively.
To do so, we measure the \ion{He}{1} \W4471 and \ion{He}{2} \W4686 luminosities, $L_{4471}$ and $L_{4686}$,
from the SDSS spectra and reverse Equations~5 \& 6 from \citet{schaerer98} such that
\begin{equation}
    Q_1 = 4.50\times10^{12} L_{4471} / f_{\gamma}	
\end{equation}
and     
\begin{equation}
    Q_2 = 9.80\times10^{11} L_{4686} / f_{\gamma}.
\end{equation}
The calculated minimum \ion{H}{1}, \ion{He}{1}, and \ion{He}{2} ionizing photons produced 
($f_{\gamma} = 1.0$) suggest that $\gtrsim10\%$ of the ionizing photons have energies large enough
to ionize He$^0$ and another $\gtrsim1\%$ are large enough to ionize He$^+$ (reported in Table~1). \looseness=-2

Observations suggest that the universe became reionized when galaxies produced ionizing 
photons at the rate of log($\xi_{\rm ion}/{\rm Hz\, erg}^{-1}) = 25.2$ photons assuming an 
escape fraction of $f_{esc}=0.2$ \citep{robertson13}. 
Adopting this $f_{esc} = 0.2$ ($f_\gamma = 0.8$), we recalculate $Q_0$.
The corresponding log $\xi_{\rm ion}^0 = 25.23, 25.28$ values for our pair of galaxies (J104457, J141851) 
are entirely consistent with galaxies that could reionize the universe.
If we also consider the $Q_1$ and $Q_2$ ionizing photons, we find 
log $\xi_{\rm ion}^{0,1,2} = 25.26, 25.31$ for our galaxies, increasing the total number of ionizing photons by $\sim8\%$.
Further, since the $Q_1$ and $Q_2$ ionizing photons are significantly less likely to be absorbed
by neutral H or He gas, and a single He$^{+2}$ recombination can result in two H$^0$ ionizations 
\citep{osterbrock89},
He ionizing photons also play a role in the cosmic reionization of hydrogen.

\subsection{Predicting LyC Escape} 
The large $\xi_{\rm ion}$ values determined for our galaxies are unsurprising given 
that H-ionizing photon production efficiency has been found to correlate with high 
ionization, low metallicity, and blue UV spectral slopes \citep{shivaei18}.
Intense high-ionization emission lines in local galaxies have also been found 
to correlate with $\xi_{\rm ion}$ (e.g., large [\ion{O}{3}]\W\W$4959+5007$ EW, 
\citealp{chevallard18,tang18}), and suggest that short, powerful bursts of star 
formation in metal-poor, low-mass galaxies could provide enough H-ionizing photons 
to achieve the log $\xi_{\rm ion} = 25.2$ value \citep{robertson13} needed to maintain 
ionization of the IGM \citep{chevallard18}.
Relative to these works, the two galaxies presented here seem to have H-ionizing 
production efficiencies typical of low-mass, metal-poor, high-ionization $z\sim0$ 
galaxies (log $\xi_{\rm ion}^0 \geq 25.13, 25.18$, depending on $f_{esc}$). 
However, the empirical relationship between $\xi_{\rm ion}$ and total 
EW([\ion{O}{3}]\W\W$4959+5007$) from \citet{chevallard18} predicts log $\xi_{\rm ion} = 25.77$ 
and $25.76$ for the extreme [\ion{O}{3}] emission observed in our galaxies 
(EW([\ion{O}{3}]) $= -1487$ and $-1455$ \AA, respectively). 
Similarly, \citet{schaerer18} find a trend of increasing $\xi_{\rm ion}$ with \ion{C}{3}] 
\W\W1907,1909 EW and note a large \ion{C}{3}] EW for the local leaker, J1154+2443, which 
has a large LyC escape of $f_{esc}=0.46$.
Extrapolating this \ion{C}{3}] EW trend predicts log $\xi_{\rm ion} > 25.8$ 
for our galaxies (EW(\ion{C}{3}]) $= -16.4$ and $-18.4$ \AA). \looseness=-2

The very large $\xi_{\rm ion}$ values predicted from the [\ion{O}{3}] and \ion{C}{3}] 
EWs are roughly 4 times larger than the canonical value, implying escape fractions
of $\sim70-80$\%.
Such strong H-ionizing production rates, and the subsequently inferred escape fractions,
are certainly atypical, but not unprecedented. 
\citet{schaerer16,schaerer18} report observed log $\xi_{\rm ion}$ values between 
25.5-26.2 for 9 local leakers with $f_{esc}=0.06-0.72$, 
that reduce down to log $\xi_{\rm ion}=25.1-25.9$ when corrected for dust attenuation.
Strong production (log $\xi_{\rm ion}>25.7$) has also been reported
for 5 nearby galaxies in \citet{chevallard18} and 2 $z\sim2$ galaxies in \citet{tang18}.

\subsection{Additional Factors Affecting LyC Escape}
Probably the most significant and limiting uncertainty in predicting the escape 
of ionizing photons is the effect of dust.
First, \citet{shivaei18} found that differences in the assumed UV reddening curve 
can result in discrepant $\xi_{\rm ion}$ estimates by as much as 0.4 dex. 
Second, none of the available reddening laws have been calibrated to the very low metallicities  
considered here, nor do their wavelength coverage extend to the He ionizing continuum regime.
In other words, the fate of the He ionizing photons due to dust is completely unknown.

A second consideration is the distribution of absorbing gas surrounding our galaxies.
Some insight can be gained from the strong absorption profiles of low-ionization 
species seen in the COS spectra.
The velocity profiles for the \ion{Al}{2} \W1671 (5.9858 eV) and \ion{Si}{2} \W1526 
(8.1517 eV) absorption features are plotted in the right column of Figure~3.
The deep profiles indicate that the gas in these galaxies are likely optically thick to 
low-ionization emission, with Si covering fractions of roughly 90\% and 70\% for 
J104457 and J141851, respectively.
Using the empirical relationship from \citet{chisholm18a}, these covering fractions correlate
to escape fractions of $\lesssim10$\% for ionizing photons with energies near $13.6$ eV. \looseness=-2

\subsection{Future Modeling}
While the arguments presented in this work for \ion{C}{4} emission as an indicator
for escaping ionizing photons may be qualitatively compelling,
radiation transfer models are necessary to draw more quantitative conclusions.
Further, it would be advantageous to resolve the profiles of multiple resonant emission lines,
such as both Ly$\alpha$ and \ion{C}{4}, in order to investigate radiative transfer through 
both low- and high-ionization gas.
Unfortunately, no high-resolution observations of both nebular Ly$\alpha$ and
\ion{C}{4} emission within the same galaxy currently exist.
Therefore, follow-up observations are needed to compare \ion{C}{4} with Ly$\alpha$
within the same galaxies and constrain models of the escape fraction of both 
high- and low-energy ionizing photons in order to determine a total escape fraction over all energies. 
This total escape fraction will help inform what photon energies escape 
reionization-era systems and whether a sufficient number of ionizing photons escape 
from star-forming galaxies to reionize the universe.

\section{Summary}
We present high-resolution {\it HST}/COS spectra of two nearby extreme UV emission-line 
galaxies, J104457 and J141851, with properties that liken them to reionization-era systems.   
We report the first observations of narrowly-separated, double-peaked, resonantly-scattered 
\ion{C}{4} emission-line profiles and intense nebular \ion{He}{2} emission.
While the low-ionization absorption line profiles of these galaxies
suggest that they are optically thick to \ion{H}{1} ionizing photons, 
the strong \ion{C}{4} nebular emission and lack of ISM absorption
indicate that they have lower column densities of
high-ionization gas that may transmit high-energy photons.
Therefore, the large number of very hard ionizing photons ($E > 47.8 {\rm eV}$) 
that are produced by these galaxies and travel outward through the high-ionization media,
may similarly escape from high-redshift galaxies. 
We therefore propose that the combination of strong \ion{C}{4} and \ion{He}{2} emission
may identify galaxies, with or without escaping Ly$\alpha$, that 
produce and transmit a substantial number of high-energy photons that contribute to cosmic reionization.
How the profiles of Ly$\alpha$ and \ion{C}{4} emission relate as LyC indicators is currently unknown,
so a larger sample and future radiative transfer modeling are needed. 
Further, future rest-frame FUV spectroscopy covering \ion{C}{4} and \ion{He}{2} in galaxies 
within the epoch of reionization using the James Webb Space Telescope ({\it JWST}) or extremely 
large telescopes (ELTs) are needed to constrain the helium ionizing photons and provide an 
estimate for their contribution to the reionization budget.

\acknowledgements
We thank the referee for insightful and constructive comments that greatly
improved the discussion and interpretation of this work.
DAB is supported by the US National Science Foundation Grant AST-1715284.
We are also grateful for support for program no. 15465 provided by NASA through a grant from the Space Telescope Science Institute, which is operated by the Associations of Universities for Research in Astronomy, Inc., under NASA contract NAS 5-26555. 
 
\bibliography{mybib}{}

\clearpage

\end{document}